\documentclass[preprint2]{aastex}

\usepackage{natbib}

\def\b0{B_0}
\def\VS{V\'azquez-Semadeni}

\shorttitle{Density Spectrum}
\shortauthors{Gazol et al.}

\begin{document}


\title{Density Power Spectrum in Turbulent \\
        Thermally Bi-stable Flows}


\author{Adriana Gazol\altaffilmark{1}}
\affil{Centro de Radioastronom{\'\i}a y Astrof{\'\i}sica, UNAM, A.
P. 3-72, c.p. 58089, Morelia, Michoac\'an, M\'exico }
\email{a.gazol@crya.unam.mx}

\and

\author{Jongsoo Kim\altaffilmark{2}}
\affil{Korea Astronomy and Space Science Institute, 61-1,
Hwaam-Dong, Yuseong-Ku, Daejeon 305-348, Korea}
\email{jskim@kasi.re.kr}

\begin{abstract}
In this paper we numerically study the behavior of the density power spectrum
in turbulent thermally bistable flows. We analyze a set of five
three-dimensional simulations where turbulence is randomly driven in Fourier 
space at a fixed wave-number and with different Mach numbers $M$ 
(with respect to the warm medium) ranging from 0.2 to 4.5. The density power 
spectrum becomes shallower as $M$ increases and the same is true for the
column density power spectrum. This trend is interpreted as a consequence
of the simultaneous turbulent compressions, thermal instability
 generated density
fluctuations, and the weakening of thermal pressure force in diffuse gas. 
This behavior is consistent with the fact that observationally determined 
spectra exhibit different slopes in different regions. The values of the
spectral indexes resulting from our simulations are consistent with 
observational values. 
We do also explore the behavior of the velocity power spectrum, which 
becomes steeper as $M$ increases. The spectral index goes from a value 
much shallower than the Kolmogorov one for $M=0.2$ to a value steeper than 
the Kolmogorov one for $M=4.5$.
\end{abstract}

\keywords{hydrodynamics --- instabilities --- ISM: structure --- methods: numerical --- turbulence}


\section{Introduction}\label{sec:intro}

The turbulent nature of the interstellar medium (ISM) is believed to play a
crucial role on the determination of its density and velocity structure
(see Elmegreen \& Scalo 2004; Scalo \& Elmegreen 2004; and
Mac Low \& Klessen 2004,for recent reviews). 
A useful characterization of the properties of a random density or velocity 
distribution is given by the spatial power spectrum. This function can be 
measured for different components of the interstellar gas by using different 
observational techniques. In the warm ionized medium (WIM) interstellar
scintillation (ISS) 
measurements have lead to a Kolmogorov slope density power spectrum, which is
valid for scales expanding over $\approx 5$ orders of magnitude 
from $10^{8.3}$ to $10^{15}$ cm (e.g. Armstrong, Rickett \& Spangler 1995).
The same authors found that when ISS data are combined  with measurements
of differential Faraday rotation angle and electron density gradients,
constraints can be put on the spectrum at much larger scales, suggesting
an extension of the precedent result to scales of $\approx 10^{20}$cm.   

For the atomic gas, the 21cm hydrogen line provides 
a variety of useful tools to measure statistical properties of turbulence. 
In particular, some estimates of the spatial power spectrum for 
the HI 21cm emission distribution in the Galactic Disk show a power
law  behavior  with different slopes.  From individual channel maps,
Crovisier \& Dickey (1983), reported a value of $-3$\footnote{Note that this
is an average value and that the slopes of individual channels vary from
$\approx -2$ to $\approx -4$} while Green  (1993) found a range from -3 to 
-2.2. For the Small Magellanic Cloud (SMC)
Stanimirovi\'c et al. (1999) also report a power law behavior
with an averaged slope of $-3.04\pm 0.02$, for the whole galaxy and an averaged
slope of $-2.85\pm 0.02$ for the eastern Wing. 
A similar behavior has been reported by Elmegreen, Kim \&
Staveley-Smith (2001) for the HI integrated emission of
the central region of the Large Magellanic Cloud. 
The same authors found a change in the spectrum slope, flattening
from -3.66 to -2.66, at $\approx 100$pc and interpreted it as the
consequence of scales in the plane of the sky becoming larger than
the depth of the medium. In that case a two dimensional medium
would be described at  scales larger than $\approx 100$pc.  
A  recent study of the HI intensity fluctuations 
power spectrum of six  dwarf galaxies (Dutta et al. 2009) reports shallower 
power law behaviors with slopes of $\approx -1.5$ for one group
of galaxies and  $\approx -2.6$ for another. They interpret the
shallower spectrum of the first group as possibly due to the presence of
two-dimensional (2D) turbulence. 
 
The observed spatial power spectrum contains information
about both, the three-dimensional (3D) density distribution and the 3D
 velocity field, and its inversion 
to a unique 3D density  or velocity power spectra is not easy. 
A method to relate the fluctuations of intensity in position-position-velocity
data cubes by changing the thickness of the velocity slice has been proposed 
by Lazarian \& Pogosyan (2000).
This method, called Velocity Channel Analysis (VCA), has been  numerically proved for high resolution isothermal 
simulations by (Padoan et al. 2006) and applied in statistical studies 
of galactic and extragalactic atomic gas.
Stanimirovi\'c \& Lazarian (2001) analyzed the spectrum of the observed
intensity fluctuations as a function of the velocity slice thickness
in the SMC and  obtain spectral indices 
of -3.3 and -3.4 for the 3D density and the 3D velocity spectra, respectively. 
For the Milky Way, two regions in the fourth Galactic quadrant have
been studied by Dickey et al. (2001) using the same method. They find 
that for a high latitude region the slope of the 3D density spectrum is 
$-4$ with an apparently little effect of  velocity fluctuations, whereas 
for a low latitude region, the slope of the density spectrum can not
be deduced from their data, which are consistent with a velocity distribution
having a structure function with a slope of $0.2$. 
Other method to disentangle the statistical properties of the three 
dimensional cold HI  density distribution at small scales in the Perseus 
arm towards Cas A and in the outer and local arms towards Cygnus 
A has been used 
by Deshpande et al. (2000). They predict the power-law index of the 
density spectrum based on that of the observed optical depth, and obtain
a slope of $-2.75$ for the two first cases, while for the local arm towards
Cygnus A, they obtain -2.5.  
Miville-Desch\^enes et al. (2003)  presented interferometric 21 cm 
observations of Ursa Major high cirrus and used the spectra of the integrated
emission and of the centroid velocity fields to deduce the three-dimensional
spectral index of the density and velocity fields at scales of $\approx 5$pc,
and found a value of -3.6 for both.
%A1
In the context of molecular clouds other methods to measure the density power 
spectrum have been developed (see e.g. Bensch et al. 2001), and find values
in the same range as those obtained for the HI gas. For instance, Stutzki
et al. (1998) find -2.8 for the Polaris Flare. 

The interstellar density power spectrum has also been  numerically studied. 
The case of compressible isothermal flows has been addressed by Kim \& Ryu 
(2005), who found that the slope of the density power spectrum becomes 
gradually shallower as the rms Mach number ($M_{\rm rms}$) increases, 
going from a Kolmogorov slope for transonic turbulence, to -0.52 for 
$M_{\rm rms}=12$  (note that this slope corresponds to the spectrum integrated 
over the angles, in this case, the Kolmogorov slope is -5/3 instead of -11/3).
They propose that this behavior reflects the development of sheetlike and 
filamentary structures for large values of $M_{\rm rms}$ and provides a 
possible explanation for the fact that for the WIM the density power 
spectrum shows a Kolmogorov slope while the corresponding slope for the
HI tends to be shallower. 
%A2, 3
Recently, Federrath et al. (2009)
numerically studied the density statistics of high Mach number isothermal 
simulations with two different kinds of forcing, finding that for the same
Mach number compressive driving leads to a significantly steeper power spectrum
than the one obtained with solenoidal forcing.

In an attempt to understand available observations, Hodge \& Deshpande (2006)
presented a simplified model in which the complex distribution of atomic
hydrogen is determined only by the superposition of voids generated by 
supernovae, either with or without kinematical effects. They obtain a velocity
spectrum with a power law behavior close to  the Kolmogorov   spectrum
and a (column)density spectrum with a slope of -3.4 for the nonkinematical
case. For their kinematical case the density power spectrum does not develop
a range with a power law behavior.

The case of density fluctuations from magnetized isothermal turbulence
has been recently studied in detail by Kowal et al. (2007) whose results
for the power spectrum show that for subsonic turbulence, the slope
presents slight changes with the sonic Mach number, while when motions
become supersonic, the presence of pronounced small scale structure
produces a flattening of the spectra, as in the non magnetized case
(Kim \& Ryu 2005). This behavior is observed for superalfv\'enic
as well as for sublfav\'enic simulations.

The HI density distribution is however strongly influenced by the presence of 
the isobaric mode of thermal instability (TI, Field 1965) and the interplay
between this instability and turbulence has important consequences on the
formation of atomic as well as molecular interstellar structure (e.g. 
Hennebelle et al. 2009).
The density spectrum for 
cooling thermally bistable turbulent gas has been presented by Hennebelle 
\& Audit (2007), who found a -0.4 slope for the integrated spectrum.

However, the behavior of the density power spectrum of thermally 
unstable gas in different turbulent conditions and in particular its
behavior for distinct values of the sonic Mach number, has not been addressed.
Although the driving mechanisms for interstellar turbulence remain a
matter of debate (Mac Low \& Klessen 2004; Brunt et al. 2009; Federrath et al.
2010), recent observational results (Tamburro et al. 2009)
using a sample of galaxies from the HI Nearby Galaxy Survey 
show that all the studied galaxies exhibit a systematic radial decline in
the  HI line width which implies a radial decline in kinetic 
energy density of HI. Previous reports of the same behavior can be found,
for example for NGC1058 (Petric \& Rupen 2007), NGC5474 
(Rownd et al. 1994), or  NGC6946 (Boulanger \& Viallefond 1992). 

In this work we present a numerical study of the density power spectrum
for cooling turbulent gas in physical conditions similar to those in
the interstellar atomic gas. The outline of the paper is the following.
In \S \ref{sec:model} we describe the model, in \S \ref {sec:results} we 
present
results concerning the effects of varying the sonic Mach number on the density
and velocity power spectra. For the former we study the whole gas distribution
as well as the behavior of gas at specific temperature intervals. Then,
in \S \ref{sec:disc} we discuss these results in the context of
previous work and finally in \S \ref{sec:conc} we give a summary and some
conclusions. 

\section{The model}\label{sec:model}

We solve the hydrodynamic equations in three dimensions to simulate 
a cubic region of 100 pc on a side with periodic boundary conditions. 
The equations are solved using  a total variation diminishing scheme
with a linearized Riemann solver (Roe 1981; Harten 1983; Kim et al.
1999).
The energy equation includes a constant background heating rate and a 
radiative cooling function with a piece-wise power-law dependence on 
the temperature obtained by S\'anchez-Salcedo et al. (2002) as a fit of the 
standard $P$ vs. $\rho$ curve of Wolfire et al.\ (1995) 
at constant heating rate, $\Gamma_0$. In thermal equilibrium conditions 
this curve implies that the  gas is thermally unstable under the
isobaric mode for $313\;{\rm K}< T <6102\;{\rm K}$, and marginally stable
for $141\;{\rm K}< T <313\;{\rm K}$. The corresponding thermal equilibrium
densities are  $n = 0.60$, 3.2 and 7.1 cm$^{-3}$, for  $T=6102$, 313 and 141 K,
respectively. For the fit, the assumed value of the heating rate is 
$\Gamma_0= 2.51\times 10^{-26} {\rm erg s}^{-1}{\rm H}^{-1}$ 
(where H$^{-1}$ means per hydrogen atom  mass).

The random turbulent forcing is done in Fourier space at a specified narrow
wave-number band, $1 \leq k  \leq 2$, where
$k \equiv \sqrt{k_x^2 + k_y^2 + k_z^2}$.
%A5
The turbulent driving is solenoidal,
 with the Gaussian deviates having zero
mean and unitary standard deviation. The amplitude of velocity perturbations 
is fixed by a constant injection rate of kinetic energy as in the prescription 
of Mac Low (1999), with the difference that we use a distinct random 
seed at each driving time. The kinetic energy input rate is chosen as to 
approximately maintain a desired rms sonic Mach number $M$, which is expressed 
with respect to the sound speed at $10^4$ K, $c_s=9.1$ km s$^{-1}$. 
Note that dividing the box size  by $c_s$, we find a time unit of  
$t_0 = 10.8$ Myr.

In all the simulations presented in \S \ref{sec:results}, the fluid
is initially at the rest, the density and the temperature are uniform 
and have thermally unstable values ($n_0=1$ cm$^{-3}$, $T_0=2399$ K).
\section{Results}\label{sec:results}
 In this section we present results from one set of five simulations with
$512^3$ grid points and five different values of $M$ namely  0.2,
0.6, 1.3,  4.0, and 4.5 (i.e from 0.41, 1.22, 2.65, 8.16, and  9.2 when 
computed with respect to the sound speed at the initial temperature). 

\begin{figure}
\epsscale{.7}
%\plotone{ima5.eps}
\plotone{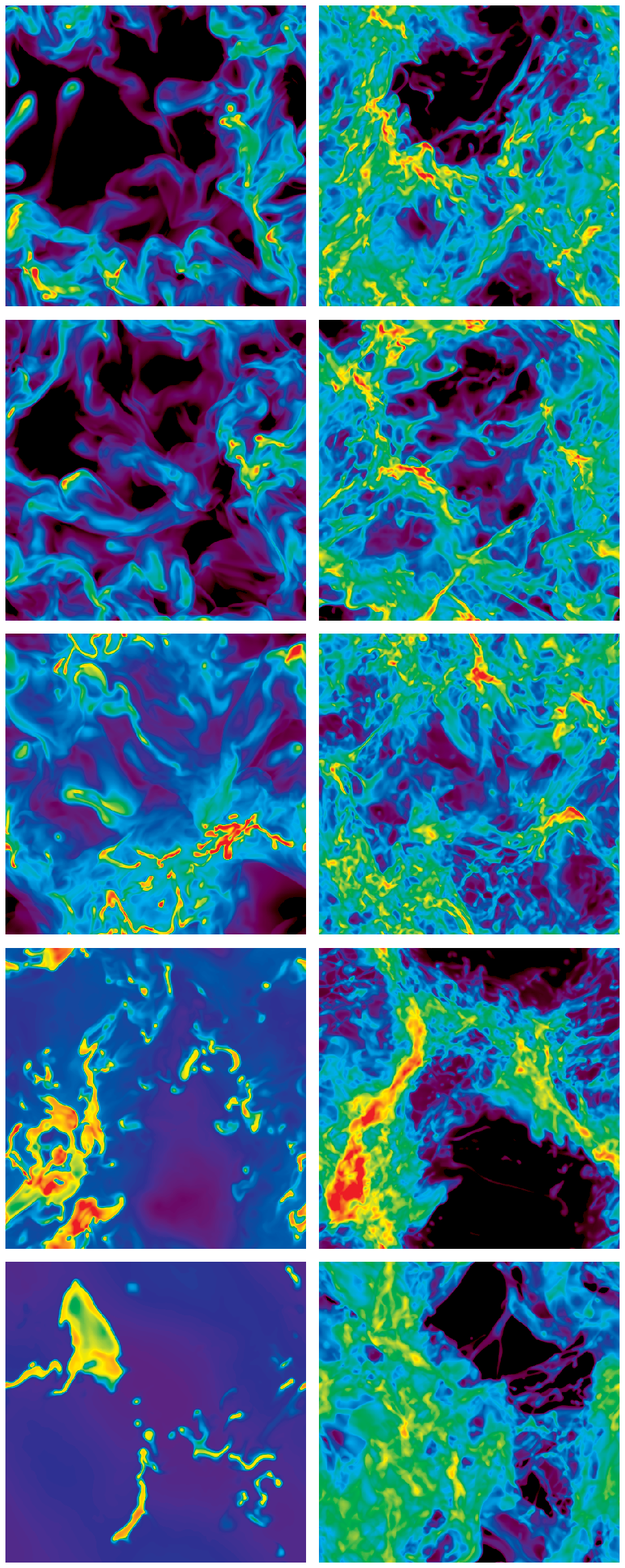}
\caption{{\it Left column~:} 
Color images of the density distribution in a two-dimensional
slice for each simulation. All the panels in this column have been 
done with the same color scale ranging from  
$10^{-1.5}$cm$^{-3}$ to $10^{1.5}$cm$^{-3}$. 
{\it Right column~:}  Column density distribution for each simulation. 
All the panels in this column have been done with the same color scale 
ranging from  $10^{-1.5}$cm$^{-3}$ and $10^{1.5}$cm$^{-3}$.
On both sides, the value of $M$ decreases from top
to bottom, from 4.5 to 0.2. Redder colors indicate denser gas.
For each simulation
the images has been done at the last time considered when computing 
averaged spectra (see Table \ref{tab:tiempos}). 
}
\label{fig:imagenes5}
\end{figure}

In Figure \ref{fig:imagenes5} ({\it left column}) 
we show two dimensional cuts of the 
density field for each simulation. All the images have been done with
the same color scale. As expected, lower $M$ simulations show a
smaller density contrast and sharper boundaries between  dense regions
and its surroundings, due to the relatively undisturbed development
of TI.  The later behavior is better appreciated in Figure
\ref{fig:imagenes2}, where different color scales have been used and 
where contours at temperatures delimiting the thermally unstable
regime have been placed. In this figure it also can be seen that for
low $M$ simulations, thermally unstable gas does just exist at the 
boundaries between gas at the two thermally stable regimes. 
\subsection{Notation}\label{sec:esp_notation}
We denote by $P_{a}(k)$ the integrated power spectrum, i.e., the integral
over spherical shells with radius $k=|{\bf k}|$ of the squared 
Fourier transform of the physical quantity $a$ and for simplicity, we refer 
to  $P_{a}(k)$ as the ``Power Spectrum''. As usual, ${\bf k}=(k_x,k_y,k_z)$
is the wave-vector.
This convention is  frequently used (e.g. Kim \& Ryu 2005, 
 Kitsionas et al. 2009) but is not universal, sometimes 
(e.g. Hennebelle \& Audit 2007) the power spectrum is defined by
the quantity $p_{a}(k)\propto k^{1-D}P_{a}(k)$, where $D$ is the dimension
of the distribution, which is the non integrated power spectrum.
 For reference, for Kolmogorov turbulence 
$p_{\bf v}(k)\propto k^{-11/3}$ whereas $P_{\bf v}(k)\propto k^{-5/3}$.

\begin{figure}
\plotone{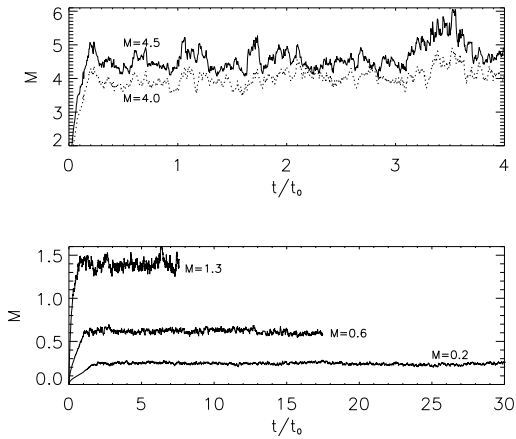}
\caption{$M_{rms}$ as a function of time for the five simulations
discused in \S \ref{sec:results}. The x-axis is code time units 
$t_0$.}
\label{fig:machrms}
\end{figure}

All the power spectra presented bellow are temporally averaged results.
The time intervals used for each simulation are listed in Table
\ref{tab:tiempos} in units of the corresponding turbulent crossing time
$t_{\rm turb}$.
%A4
We  define  $t_{\rm turb}=l_{\rm for}/(M c_s)$ with $l_{\rm for}= 50$pc being
the characteristic scale of the turbulent forcing. 
In this table we also present the final time reached by each simulation
and the number of snapshots we use to compute the average. 
% 
%A0
In Figure \ref{fig:machrms} we present the Mach number as a function of 
time for
the five simulations presented in next two sections. The x-axis has
$t_0$ units in order to emphasize that, in these units, much longer 
integration times have been used for lower values of $M$. In $t_0$ units,
the lower
limit of the time interval used for averaging power spectra (i.e. the
lower limit in the third column of Table \ref{tab:tiempos}) corresponds
to 10.0, 5.0, 2.3, 1.0, and 1.0 for $M$ equal to 0.2, 0.6, 1.3,  4.0, and 4.5,
respectively. It can thus be seen from the figure that the power spectra 
have been averaged in a stationary regime.

\subsection{Density and Velocity Spectra for the whole simulation}\label{sec:esp_completos}
\begin{figure}
%\plotone{esp_ro_compfittt.eps}
\plotone{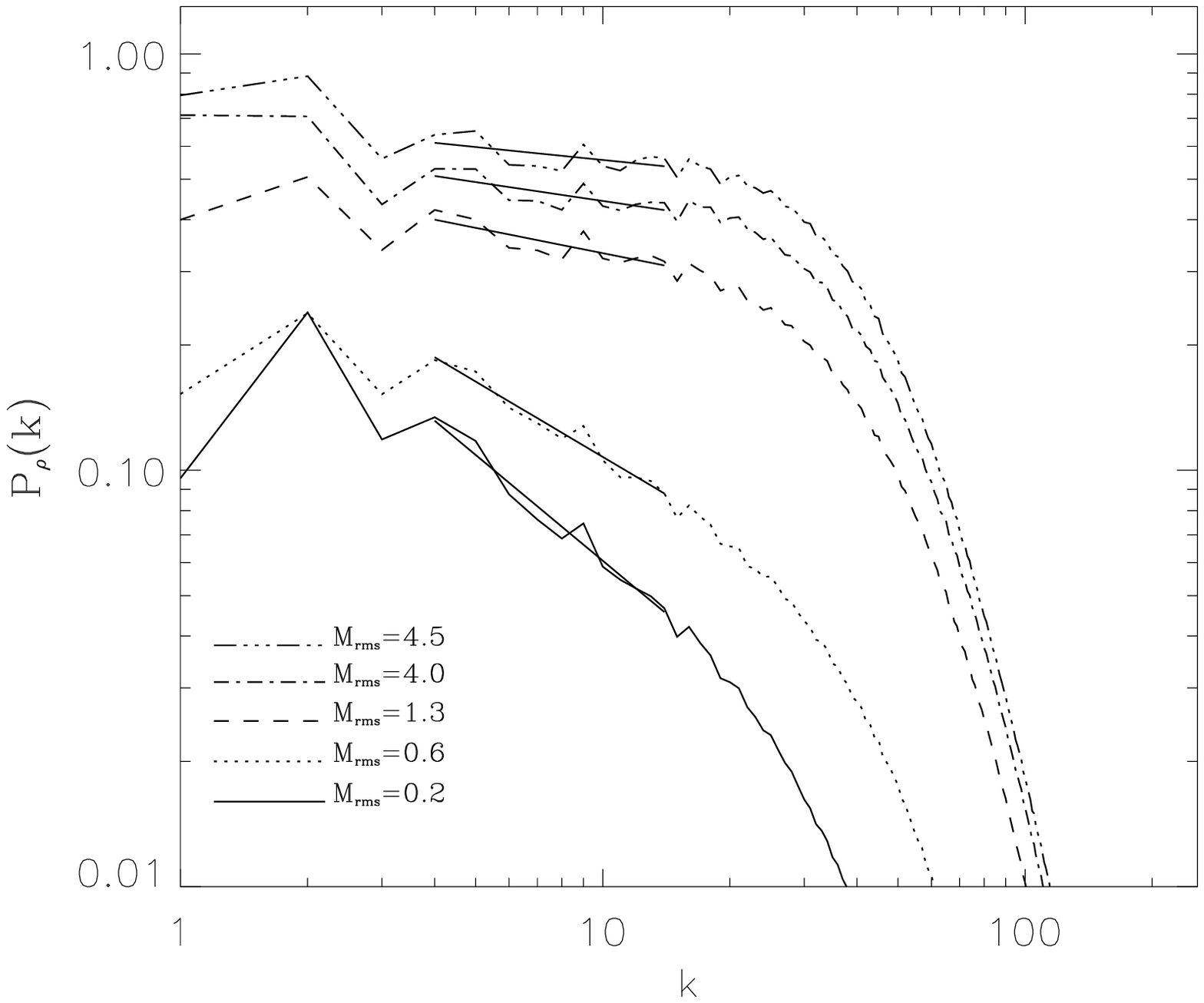}
\caption{Density power spectra $P_{\rho}$ for simulations with
different $M$. Solid lines represent least-squares fits over the
range $4 \leq k \leq 14$.}
\label{fig:espectros_den}
\end{figure}

\begin{figure}
%\plotone{esp_rocol_compfittt.eps}
\plotone{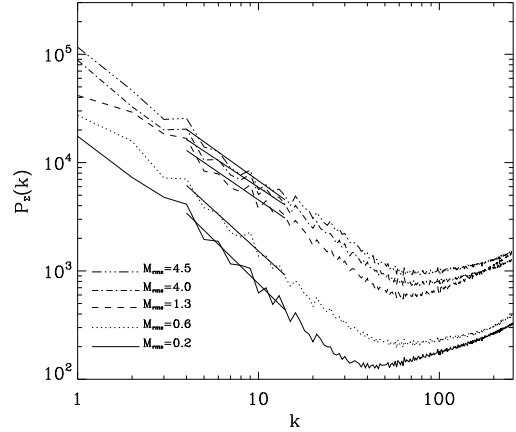}
\caption{Column density power spectra $P_{\Sigma}$ for simulations with
different $M$. The two dimensional density distribution is obtained 
by projecting the full three-dimensional density field on the $z$ direction.
Solid lines represent least-squares fits over the
range $4 \leq k \leq 14$.}
\label{fig:espectros_dencol}
\end{figure}

Time averaged density power spectra $P_{\rho}$, are shown in Figure 
\ref{fig:espectros_den} along
with power law least-squares fits computed for $4\leq k \leq 14$,
which implies physical scales ranging from $\approx$7pc to 25 pc. 
The power law indexes obtained from those fits for each simulation
are listed in Table \ref{tab:indices} along with power law indexes
resulting from similar fits for all the power spectra discussed in the 
present paper. 
As in the isothermal case reported by  Kim \& Ryu (2005), the absolute value 
of the slope (in logarithmic coordinates) decreases as the value of $M$ 
increases. For our simulations it goes from $-0.84$ for $M=0.15$ to $-0.10$ 
for $M=4.5$. These values are consistent with the value previously
reported by Hennebelle \& Audit (2007). We have also computed 
column density power spectra $P_{\Sigma}$, by projecting the 3D density field 
on the $z$ direction  and using the resulting 2d distribution. 
As expected, due to the reduced amount of points, these spectra 
(displayed in Figure \ref{fig:espectros_dencol}) develop more fluctuations 
than those obtained from the full three dimensional distribution.
 The best power law fits of these spectra, whose slopes are listed
in  Table \ref{tab:indices},  show that they
are steeper than spectra obtained from the full 3D distribution and  have a 
similar trend as them, namely that they become flatter as $M$ is increased,
with values ranging from $-1.64$ for $M=0.2$ to  $-1.11$ for $M=4.0$. Note 
however that the slope of the column density power spectrum for  $M=4.5$ is
larger than the one for $M=4.0$. As expected, the power-law
region of $k^{-1}P_{\Sigma}$ behaves in approximately the same
way as  the power-law region
of  $k^{-2}P_{\rho}$ implying that $p_{\Sigma}$ and $p_{\rho}$ have
approximately the same spectral index as predicted by Stutzki et.al. (1998) 
and shown by Brunt et al. (2010). 
%A6
This fact has previously been numerically confirmed by 
Mac Low \& Ossenkopf (2000) and Federrath et al. (2009).  
For all the spectra in Figure 
\ref{fig:espectros_dencol} it can also be seen a change of sign
in slope  at high frequencies. This change, that does not occur for
$P_{\rho}$, is due to the presence of small three dimensional structures that 
do not overlap when projection is done (see Figure \ref{fig:imagenes5}, {\it right column}).  

\begin{figure}
\epsscale{1.}
%\plotone{esp_v_compfittt.eps}
\plotone{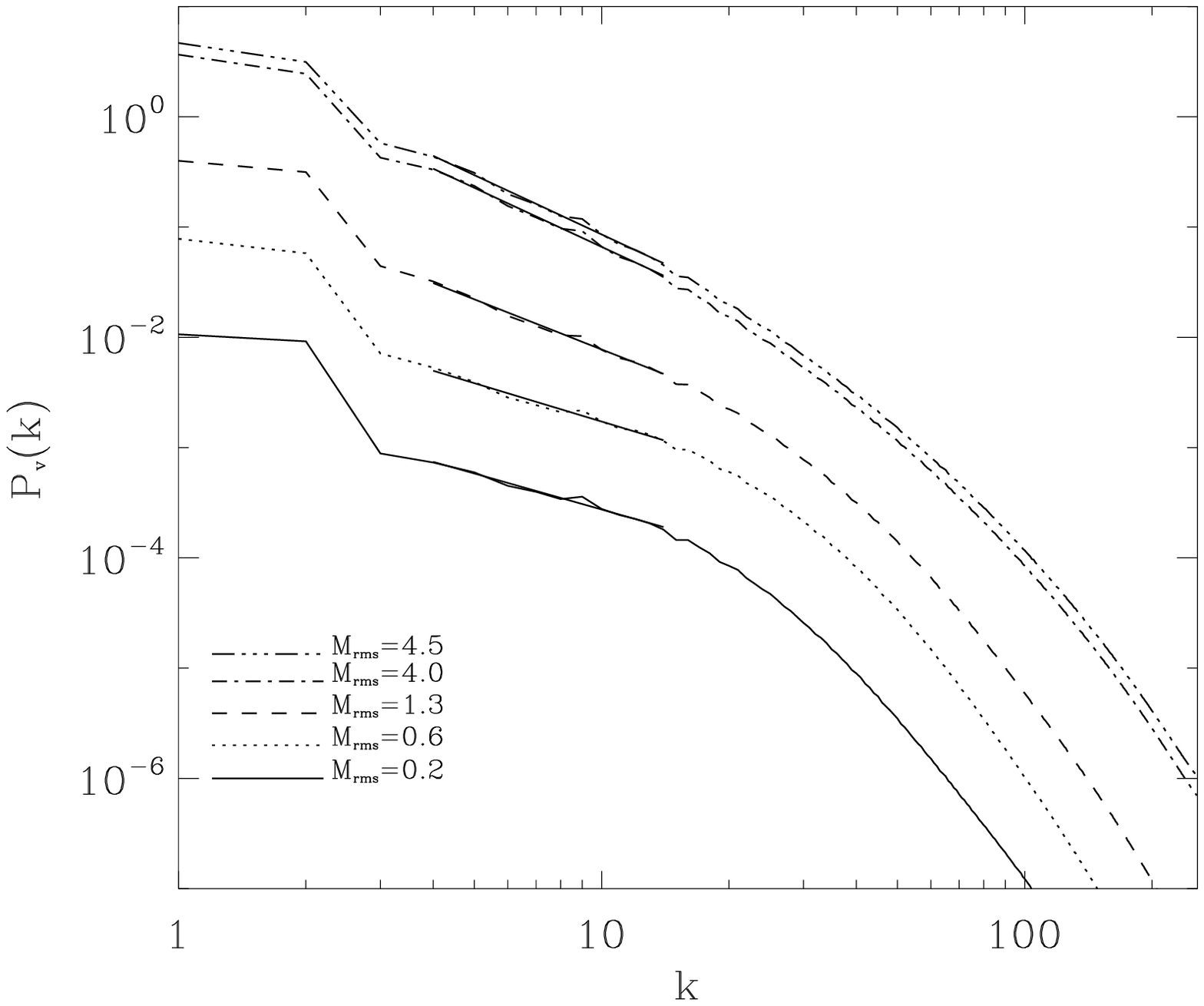}
\caption{Velocity power spectra $P_{v}$ for simulations with
different $M$. Solid lines represent least-squares fits over the
range $4 \leq k \leq 14$.}
\label{fig:espectros_vel}
\end{figure}

It is also interesting to study the behavior of the velocity power spectrum 
$P_{\bf v}$
as $M$ changes (see Fig. \ref{fig:espectros_vel}), which shows significant
variations when the Mach number changes. In particular, the slope resulting
from a power-law fit at the same wave number range as in the density 
spectra case,  goes from
-1.79 (steeper than the Kolmogorov slope of -5/3) for $M=4.5$ to $-1.08$ 
for $M=0.2$.
This behavior is different from the one reported by Kim \& Ryu
(2005), for the one dimensional case, where the slope was nearly equal $-2$
irrespective of the rms Mach number. A similar slope has been reported by
Kritsuk et al. (2007) for highly compressible isothermal turbulence. 
%A7
Recently, for an isothermal gas with $M=5.5$ Federrath et al. (2009) 
found slopes of $-1.86\pm 0.05$ and  $-1.94\pm 0.05$ for solenoidal and 
compressive forcing, respectively; whereas
Schmidt et al.  (2009) obtained a value of $-1.9$ for $M=2.5$, also in
the isothermal case. 
For two-dimensional non isothermal gas with $M\approx 1$, 
Hennebelle \& Audit (2007)
obtain a velocity spectrum in good agreement with the Komogorov law. 

\begin{figure}
%\plotone{esp_vro_compfittt.eps}
\plottwo{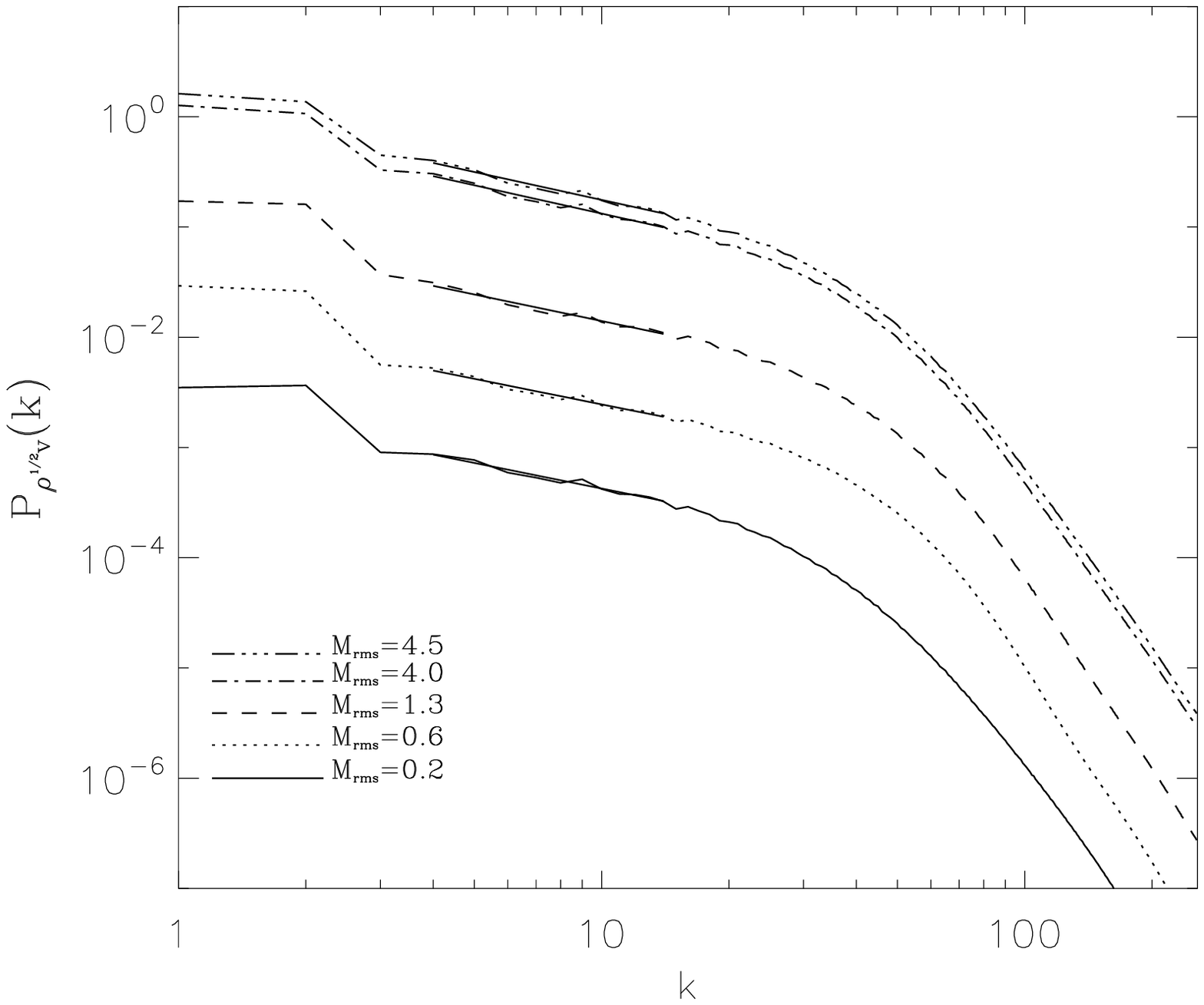}{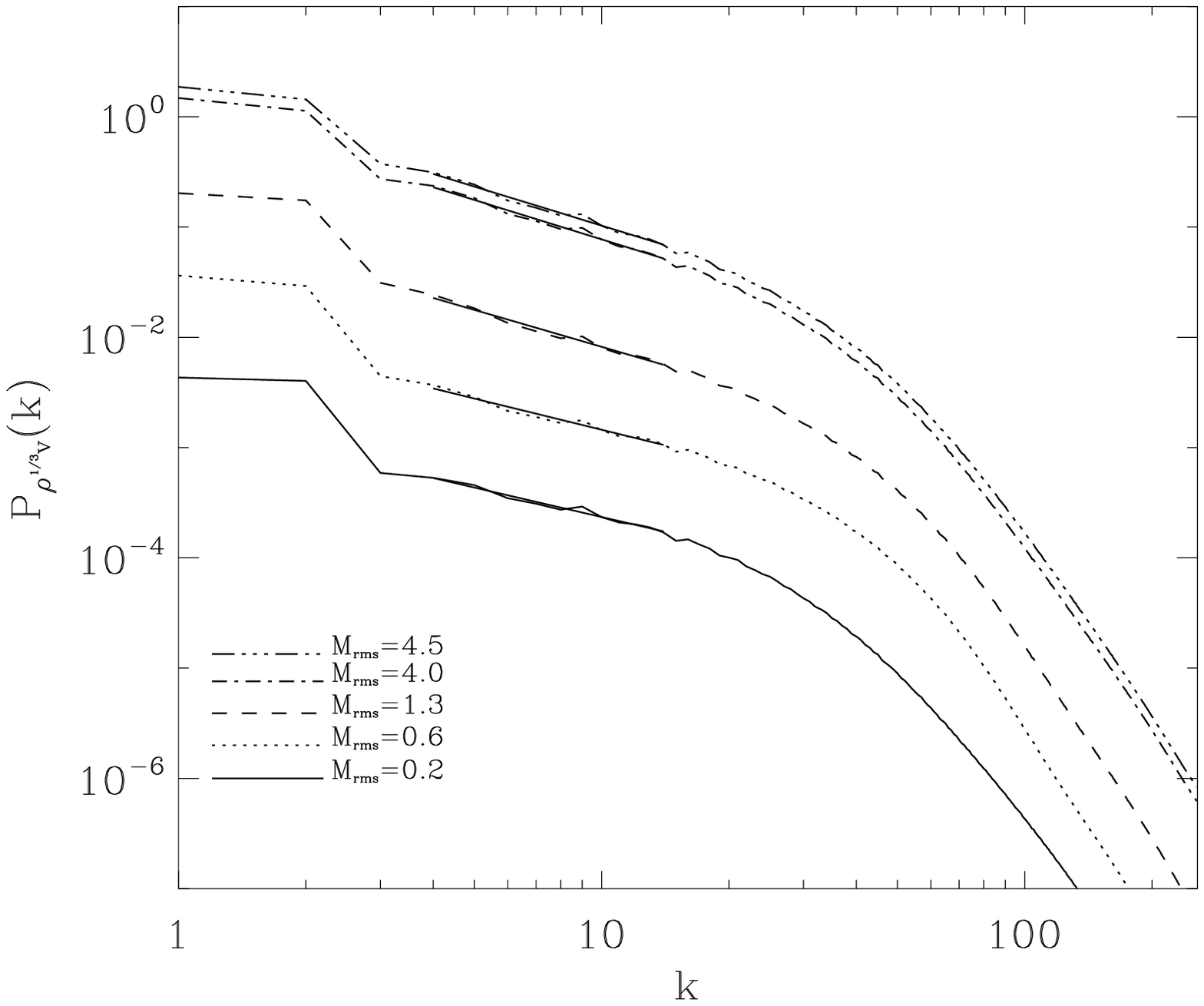}
\caption{Power spectra of ${\rho}^{1/2}v$, $P_{{\rho}^{1/2}v}$ ({\it left})
and ${\rho}^{1/3}v$, $P_{{\rho}^{1/3}v}$ ({\it right})
for 
simulations with different $M$. 
Solid lines represent least-squares fits over the
range $4 \leq k \leq 14$.}
\label{fig:espectros_velden}
\end{figure}

%A8    
As stated by Kitsionas et al. (2009) the density weighted spectrum 
represents an appropriate statistical tool to study compressible 
turbulence because it takes into account the fact that in the 
compressible case most of the mass ends up in small volumes. 
This spectrum however can be computed as the power spectrum
of the quantity ${\rho}^{1/2}{\bf v}$ which is directly related to the 
kinetic energy, or as the power spectrum of ${\rho}^{1/3}{\bf v}$, 
that arises from the condition of constant mean volume energy transfer rate, 
which is valid for a compressible fluid in a steady state 
(Lighthill 1955, see also Kritsuk et al. 2007). Here we show both.  
In Figure \ref{fig:espectros_velden}a
the power spectrum of $\sqrt{\rho}{\bf v}$ is shown. As expected due to the
increased density power at small scales, these spectra are  notoriously flatter
than  $P_{\bf v}$ with power-law fit slopes ranging from $-0.77$ 
to $-0.85$. Moreover, in this case the slope value range is narrower than
for the density and the velocity cases and  do not show any systematic
behavior when varying $M$. The slopes we get are comparable to the value
reported by Hennebelle \& Audit (2007) for two dimensional simulations.
On the other hand, $P_{{\rho}^{1/3}{\bf v}}$ 
(Fig. \ref{fig:espectros_velden}b), is steeper than $P_{{\rho}^{1/2}{\bf v}}$, 
and except for the lower value of $M$ follows the same trend as the velocity
spectrum, i.e. it steepens as $M$ decreases.  The last fact can be understood 
as a consequence of the lower dependence on $\rho$. Also, the slope value range
is larger for $P_{{\rho}^{1/3}{\bf v}}$ but is still narrower than for
the density and the velocity spectra. 
This implies that for our simulations the quantity $\rho^{1/2}v$ has
more well defined scaling law than the quantity $\rho^{1/3}v$.

\subsection{Density Spectra for Different Temperature Ranges}
\begin{figure}
\epsscale{.8}
%\plotone{esp_rofases_multiplot.eps}
 \plotone{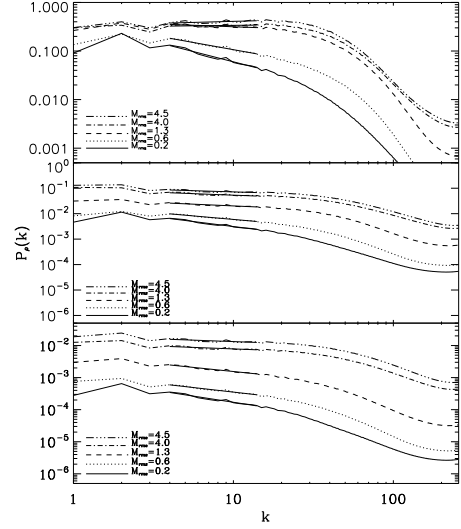}
\caption{Density power spectra for gas at different temperature
ranges resulting  from simulations with
different $M$. In the three panels solid lines represent 
least-squares fits over the range $4 \leq k \leq 14$. 
{\it Top~:} Density power spectra for cold gas ($T\leq 313$K).
{\it Middle~:} Density power spectra for thermally unstable 
gas ($313 {\rm K}< T < 6110$K).
{\it Bottom~:} Density power spectra for warm gas ($T\geq 6110$K).   
}
\label{fig:espectros_rofases}
\end{figure}

The pure development of the isothermal mode of thermal instability (TI) in
the atomic galactic gas allows to distinguish three thermodynamic regimes:
stable warm gas, unstable gas, and stable cold gas.
Figure \ref{fig:espectros_rofases} shows the density power spectra we 
obtain for each of these regimes. 
The temperature limits we use to separate the gas are arbitrarily 
chosen as those determined by the cooling function, i.e.
313 K and 6110 K, implying that our warm, unstable and cold gases
have temperatures, $T>6110$K, $313{\rm K}<T<6110$K, and $T<313$K, respectively.
 As a first approximation we compute each spectrum by considering only
those grid points where the temperature lies in the desired range and
assigning a zero value elsewhere. We are aware that this procedure
can introduce high frequency noise in the spectrum but this effect
should affect in a comparable way all the spectra in a given temperature
range. Also we recall that the slopes are measured for relatively small
values of $k$ (see \S 3.2), which is hardly affected by the high frequency 
noise.
 
%The density power spectra for warm gas $P_{w\rho}$, the unstable
%gas $P_{u\rho}$ and the cold gas 
%$P_{c\rho}$ follows approximately the same behavior as $P_{\rho}$ namely
%it becomes flatter as $M$ increases. 
%Note that for the warm and the cold gas this is not satisfied for 
%the largest value of $M$
%and that  for large values of $M$ the cold gas spectrum has a 
%positive slope.

 The density power spectrum for the warm gas $P_{w\rho}$, the unstable
gas $P_{u\rho}$ and the cold gas 
$P_{c\rho}$ follows  the same behavior as $P_{\rho}$, namely
it becomes flatter as $M$ increases (see Table \ref{tab:indices}). 
When comparing the slopes obtained for spectra at different
temperature regimes for the same  value of $M$, the sole consistent
behavior is that at large $M$ (transonic or supersonic with respect
to the gas at $10^4$K) the spectrum flattens as the temperature decreases,
and for $M=4.0$ and $4.5$ the slope  even changes its sign. 
As the averaged turbulent velocity is approximately the same in all the three
temperature ranges implying that colder gas has considerably larger Mach 
numbers, this behavior is not surprising because is the same as the one 
reported by  Kim \& Ryu (2005). The slopes are however much shallower than the 
ones they obtain. For instance, in our simulation with $M=1.3$ the slope of
the power spectrum for cold gas is $-0.12$, whereas the slope of the
power spectrum they get for an isothermal simulation with $M=7.3$ is 
$-0.75$ (note that the Mach number for gas at $313$K corresponding to 
$M=1.3$ is 7.35).
 
\subsection{A comparison with the isothermal case}\label{sec:isot}

\begin{figure}
\epsscale{1.}
%\plotone{esp_ro_res.eps}
\plottwo{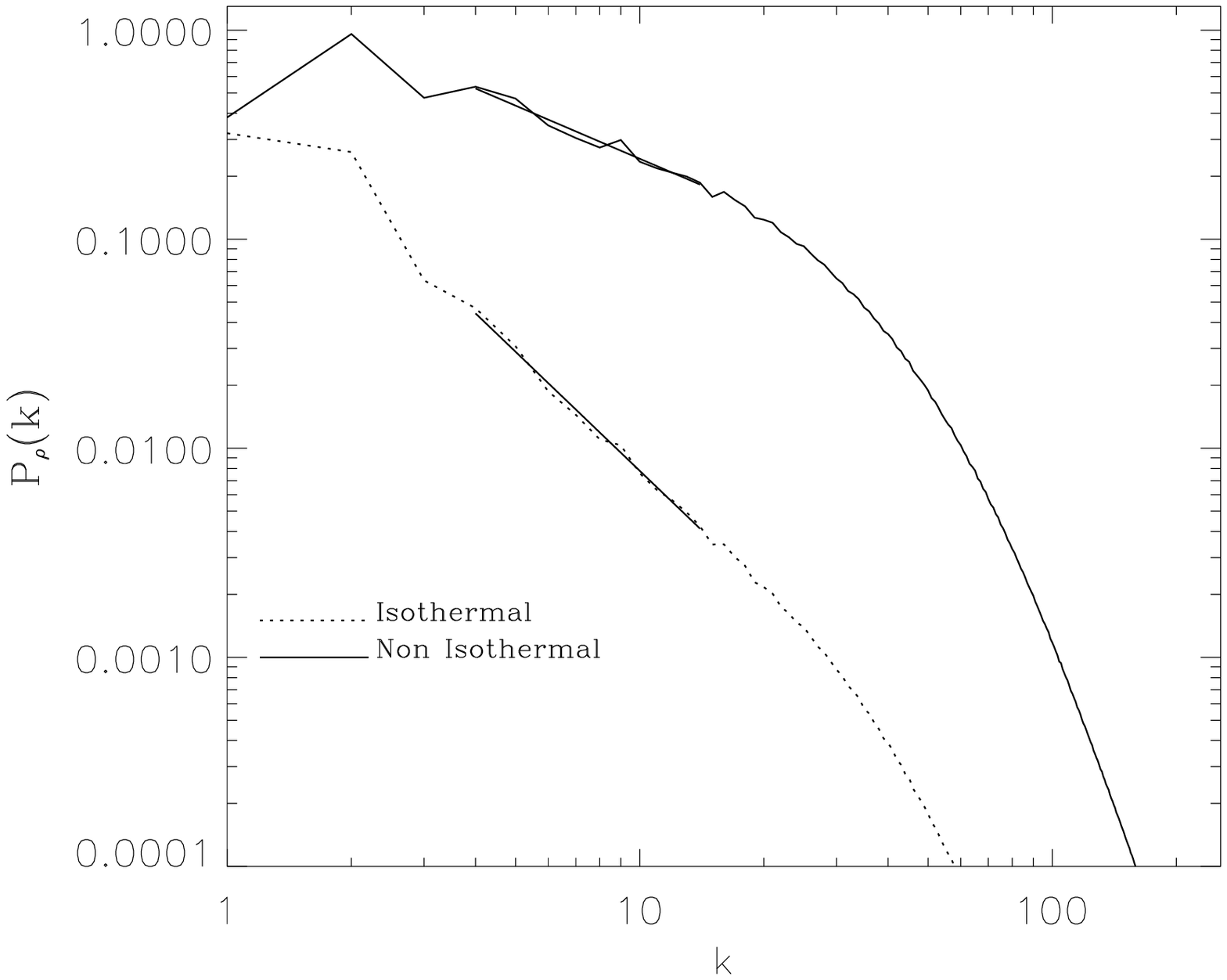}{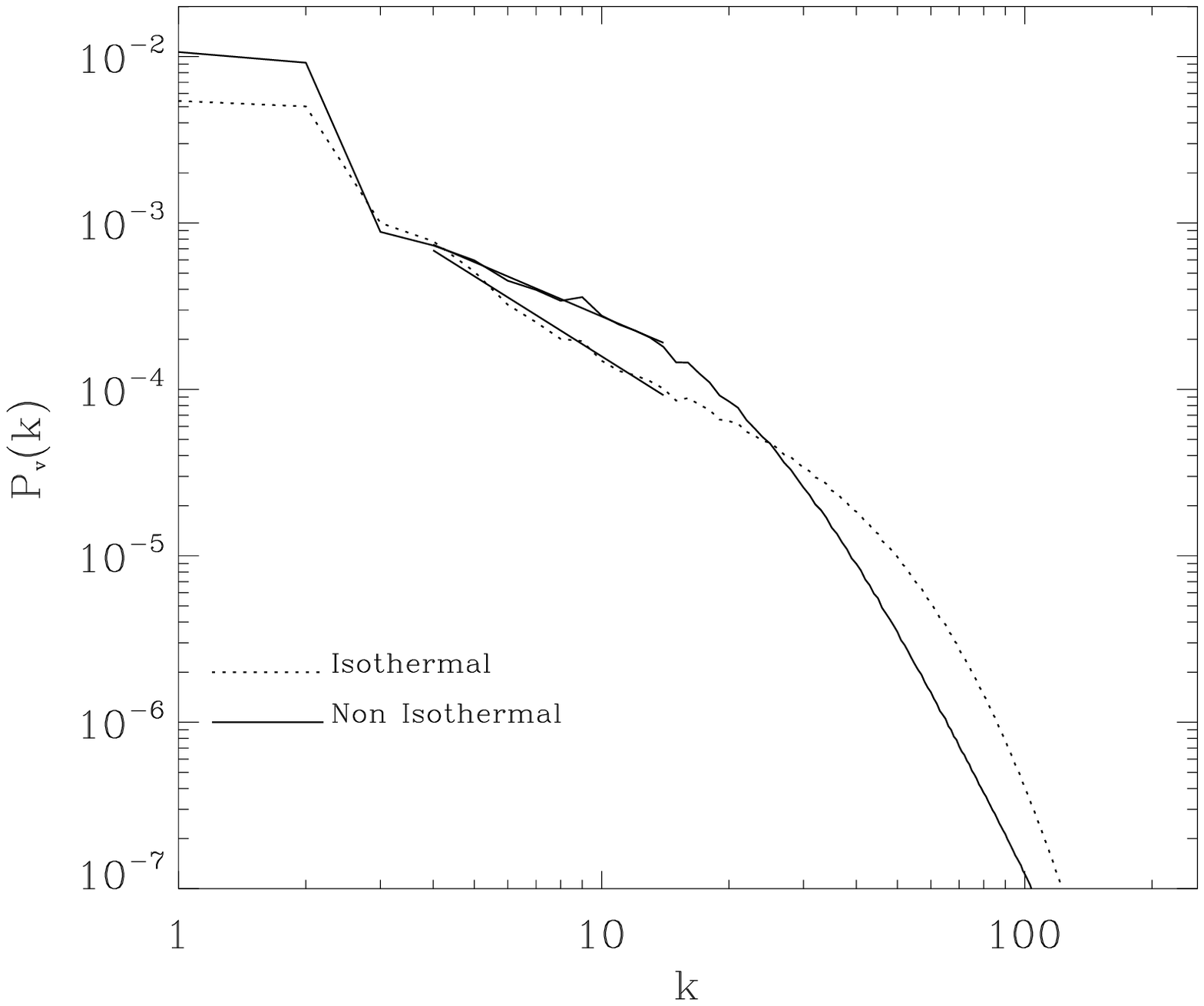}
\caption{Density ({\it left}) and velocity ({\it right}) power spectra for
a thermally unstable simulation ({\it solid  line}) and an isothermal
one ({\it dotted  line}), both at $M=0.2$. 
Solid lines represent least-squares fits over the
range $4 \leq k \leq 14$.}
\label{fig:compiso}
\end{figure}

In order to confirm that turbulence statistics is significantly affected
by the presence of TI, in this section we compare a thermally unstable
simulation with an isothermal one. We use our lower $M$ simulation because
in this case the velocity power spectrum for the isothermal regime is expected
to have a slope close to that predicted by Kolmogorov (i.e. -5/3). For 
completeness sake we do also include a comparison between density power
spectra. In Figure \ref{fig:compiso} we show density ({\it left}) and
velocity spectra ({\it right}) for the thermally unstable ({\it solid line})
and the isothermal ({\it dotted line}) simulations. For the former case
the fitted slopes are the same as those reported in Table \ref{tab:indices}, 
while for the isothermal case slopes are -1.89 and -1.60 for the density
and the velocity spectrum, respectively. The first value is consistent
with the trend found by Kim \& Ryu (2005) and the second one is close to the
expected value. These slopes  indicate that the effects presented in 
previous sections concerning the flattening of the density and the velocity 
power spectra in thermally unstable flows are not due to the time
interval we use to compute averaged values and that the presence of TI has
in fact important effects on turbulence statistics.
Note that all the spectra in Figure \ref{fig:compiso}  have been computed 
by averaging values in the same time interval (see Table \ref{tab:tiempos})
and that the isothermal simulation has been done with an isothermal version 
of the same code as the one described in \S \ref{sec:model}.

\section{Discussion}\label{sec:disc}

\subsection{Relation with Previous Work}\label{sec:other}
%A9
Observational results on the HI  power spectrum show 
a power law behavior with  an extended range of slope values.
This is true for the spectrum resulting from intensity fluctuations  
as well as  for the spectrum deduced for the underlying 3D density 
distribution.
The latter spectrum describes the statistics of a density distribution
strongly influenced by the presence of TI and turbulence. 
%A10
On the other hand, it is
well known that TI can be triggered in the warm stable gas by turbulent
motions characterized by large enough sonic  Mach numbers 
(Hennebelle \& P\'erault 1999) and that the density PDF is
strongly influenced by the characteristics of turbulence (Klessen 2000;
Federrath et al. 2008)
as well as by the equation of state (e.g. Passot \& V\'azquez-Semadeni
1998, Nordlund \& Padoan 1999, Li et al. 2003, Gazol et al. 2005, 
Kowal et al. 2007). In particular Passot \& 
V\'azquez-Semadeni (1998) showed that the density PDF is lognormal for 
isothermal flows ($\gamma=1$, with $\gamma$ being the effective
polytropic index), but develops a power-law tail at
high densities for $\gamma < 1$, and at low densities for $\gamma >1$.
 For a polytropic self-gravitating gas  Li et al. (2003)
found imperfect lognormal distributions whose width increases as $\gamma$
increases.
The case of thermally unstable gas interacting with turbulent motions
have been studied by Gazol et al. (2005). Using
the same cooling function as the one used for the present work, they
found bimodal distributions that become broader  as $M$ or the
forcing scale increases. The bimodal character of the distributions
is less pronounced for small forcing scales and for large values of
$M$. 
On the other hand, different turbulent conditions have been suggested
by the observational estimation of the available kinetic energy 
in the atomic gas of nearby galaxies (Tamburro et al. 2009).
In this context, the results we presented in the previous section
suggest that different slopes in the observed HI density power spectra could be
due to different turbulent conditions, in particular different Mach
numbers.
Different values of the Mach number have been used by Elmegreen et al. (2001)
in fractal models aimed to fit HI data, including the power spectrum, of 
the Large Magellanic Cloud. They found that the large amplitude of the
observed intensity variations cannot be achieved by turbulence alone; phase
transitions are required. In models presented by these authors however, both 
the Mach number and the HI scale height are simultaneously modified making 
it impossible to isolate the effect of varying $M$.

The flattening of the density power spectrum as $M$ is increased
is consistent with results reported by Kim \& Ryu (2005) for the isothermal
hydrodynamic case and by Kowal et al. (2007) for the isothermal MHD case. 
Our spectra are 
shallower than the ones they obtain but they are consistent
with the value reported by Hennebelle \& Audit (2007) for two  dimensional
thermally bistable simulations. These authors interpret the relatively 
flat density power spectrum as a consequence of the strong and stiff density
fluctuations produced by the development of TI rather than by highly supersonic
motions. Our numerical experiments consider the effect of both. We
can understand our results in terms of the predictions of Sachiev
\& Woyczynski (1996) concerning the power spectrum of density resulting
from a velocity field governed by the Burgers equation in the presence
of pressure forces. They find that in the limit of a negligible pressure
force $P_{\rho}$ does not depend on $k$, while  in the presence of
a pressure force $P_{\rho}\propto k^{-2}$.
The effect of varying the turbulent conditions on 
the thermodynamic properties of a thermally bistable gas, has been studied
by Gazol et al. (2005). For large scale forcing they 
found that as $M$ increases, the local ratio of turbulent crossing time to 
cooling time decreases, causing transient structures in which the effective 
behavior is intermediate between the thermal-equilibrium and adiabatic regimes.
In particular, the mean pressure in a given density interval drifts away from 
the thermal equilibrium value $P_{\rm eq}$ as $M$ is increased, moving 
toward $P > P_{\rm eq}$ for the dense gas and toward $P < P_{\rm eq}$ 
for distributions centered at typical densities of the warm and the
unstable gas. Flatter spectra in high 
$M$ simulations could thus be  
produced by the weakening of pressure forces in diffuse gas. In the
simulations presented here there are then three physical ingredients 
contributing to the flattening of the density power spectrum: density
fluctuations produced by turbulence, density fluctuations produced by TI,
and 'enhanced' density fluctuations produced by the interactions between 
turbulence and TI.  

Direct comparison of the spectral indexes we obtain for $P_{\rho}$
with observations of the density power spectrum show that they only agree 
with  values reported by Deshpande et al. (2000) and with those obtained 
by Dutta et al. (2009) for nearby dwarf galaxies. 
We recall that values in Table
\ref{tab:indices} should be multiplied by a factor $k^{-2}$ in order
to be comparable with observational data (see \S \ref{sec:esp_notation}). 
Note that Dutta et al. (2009) find that, for all the galaxies
in their sample, the power-law slope remains constant as the
channel thickness is increased, suggesting that the fluctuations
in the HI intensity  are only due to density fluctuations, or that the slope of
the velocity structure function is $\approx 0$. 

Theoretical predictions by Lazarian and Pogosyan (2000)
can be used to further compare our results with observational results,
%A11
in particular with works reporting only spectral slopes from
individual channel maps or averaged values over distinct velocity channels.
The VCA  method consists on gradually increase
the velocity thickness of the sampling region  until
the slope of the observed two dimensional spectrum gets stabilized.
For the thickest slices, the velocity information is averaged out
and the power-law index of the three dimensional density spectrum $n$ 
is recovered. For thin slices  the method provides a relation between $n$ 
and the three dimensional velocity spectral index. A slice is defined 
as ``thin'' if the turbulent velocity dispersion on the studied scale 
is smaller than the velocity slice thickness. Although this method 
have not been numerically tested for thermally bistable cases
\footnote{Such a test is out of the scope of this work  but we plan to
present it elsewhere.}, we can explore what we would get if we
were observing our density distribution through a ``thin'' velocity 
slice (observing through a thick slice would give results on the second
column of Table \ref{tab:indices} plus -2). To do that, we need
to combine slopes for  $P_{\rho}$ and slopes for $P_{\bf v}$ in
the regime of a shallow three dimensional density 
( $k^{-2}P_{\rho}\propto k^{n}$ with $n>-3$) for a thin velocity 
slice (see Lazarian \& Pogosyan, 2000). 
In this regime, $n$ is related with the slope of the velocity structure 
function $m$  (the three-dimensional velocity spectrum is then
$\propto k^{-3+m/2}$) through the index of the spatial power spectrum,
which is proportional to $k^{n+m/2}$.
In that case, for the hypothetically observed intensity spectrum 
$P_{\bf obs}$, which combines information about density and velocity,  
 we obtain slopes of -2.92, -2.76, -2.71, -2.93, and -2.89 
for $M=$0.2, 0.6, 1.3, 4 and 4.5,
respectively. These values are in the range of values reported by 
Crovisier \& Dickey (1983),  Green (1993), Stanimirovi\'c et al. (1999),
and Dutta et al. (2009). In this case the behavior of $P_{\bf obs}$ as 
the Mach number
increases (i.e.  a non systematic change) is consistent with the behavior 
we find for $P_{{\rho}^{1/2}\bf v}$, which is in turn consistent with the
fact that $P_{\rho}$  and $P_{\bf v}$ show opposite trends. 
%A12
The behavior of $P_{\bf obs}$ as $M$ increases, is in contrast inconsistent 
with the behavior of  $P_{{\rho}^{1/3}\bf v}$, which tends to become 
shallower for lower Mach numbers. For our simulations, thus, 
$P_{{\rho}^{1/2}\bf v}$ seems to be a more adecuate quantity to take
into account density and velocity contributions to the statistics.
However, it should be noticed that similar slopes of 
$P_{{\rho}^{1/2}\bf v}$,  $P_{{\rho}^{1/3}\bf v}$, or  $P_{\bf obs}$ 
could result from very different physical conditions. 
%A16  
This implies that in order to relate power spectrum
properties with physical conditions in the HI gas, and in 
particular with the Mach number, the distinction between density
induced and velocity induced intensity fluctuations seems to be
unavoidable. In cases where column density maps can be directly
obtained, and the density power spectrum can be unambiguously 
measured the relationship between the density power spectrum
and the Mach number can be easily established.

On the other hand, the behavior of the velocity spectrum, steepening as
 $M$ increases, can be understood as follows.
The fact that for larger values of 
$M$ we get a slope larger than the Kolmogorov one, as other authors
have done  (e.g.  Kritsuk et al. 2007; Schmidt et al. 2009; Federrath et al. 
2010) for the isothermal 
case  is consistent with the effective polytropic index approaching 1 as 
$M$ becomes larger (Gazol et al. 2005). In high $M$ simulations
turbulent motions generated by the development of TI, which are typically
of the order of tenths of km s$^{-1}$ (Kritsuk \& Norman 2002, Piontek \&
Ostriker 2004), are much smaller than the forcing generated
velocities. Instead, at small values of $M$, the modest amplitudes of
turbulence produced by TI  can considerably contribute to
the gas velocity. Furthermore, the almost undisturbed development of
TI in small $M$ simulations occurs only at relatively small scales 
($\lesssim 10$pc, see Heitsch et al. (2008) for a detailed 
study of conditions for thermally dominated fragmentation), producing thus
a flattening of the velocity spectrum. Note that Heitsch et al. (2008) find that
outside the strictly thermally unstable regions, in particular at large scales,
there is a regime of densities, temperatures, and sizes where cooling still 
dominates and can lead to fragmentation when an external 
(ram or gravitational) pressure is applied.    
Although our results can be physically explained, they differ from the
better established observational reports 
(Stanimirovi\'c \& Lazarian 2001, Miville-Desch\^enes 2003). 
Other observational works (Dickey et al. 2001, Dutta et al. 2009)
however, present results consistent with a very shallow structure functions.
More observational data as well as a detailed high resolution numerical
study (see \S \ref{sec:res}) are thus needed in order to characterize
the velocity spectrum behavior in a thermally bistable gas.

\begin{figure}
\epsscale{.50}
%\plotone{ima2_cont.eps}
\plotone{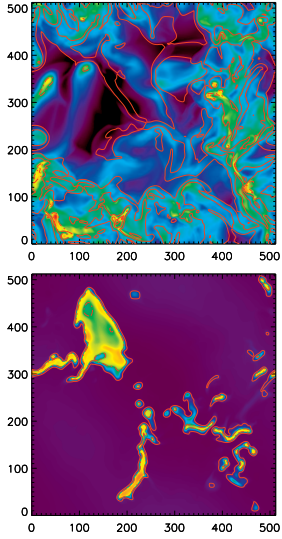}
\caption{Same density slices as in figure \ref{fig:imagenes5} for $M=4.5$
({\it top}) and $M=0.2$ ({\it bottom}) but with different color
scales. For $M=4.5$ color scale ranges from $10^{-2.5}$cm$^{-3}$ to 
$10^{2}$cm$^{-3}$ while for  $M_{rms}=0.2$ it goes from 
$10^{-1.2}$cm$^{-3}$ to $10^{1.5}$cm$^{-3}$. In both panels, 
{\it solid line} and {\it dashed line} contours are placed at $6110$K 
and $313$K, respectively.}
\label{fig:imagenes2}
\end{figure}

The behavior of the spectra obtained for specific temperature ranges
is consistent with the behavior reported by Kim \& Ryu (2005) in two
ways: when $M$ is changed within a specific range and when a fixed value of
$M$ is considered and we look to different temperature intervals.  The spectra
we get are however much shallower than those shown in Kim \& Ryu (2005).
A quantitative analysis regarding the slopes is not
possible because we do not have enough resolution to 
ensure that structures at the three temperature intervals can expand  over two
decades in scale.
The fact that a flattening of the spectrum for lower temperatures is only 
present  in turbulence 
dominated simulations is also expected because in low $M$ simulations thermal
instability can proceed with almost no disturbances, leading to a better 
segregated gas where unstable gas is present only in the thin interfaces
between cold and warm gas (see Figure \ref{fig:imagenes2}).
In turbulence dominated simulations, on the other hand, colder gas tends 
to be dominated by small scale density structure 
(see Figure \ref{fig:imagenes2}) and thermally unstable gas spread over 
larger regions. 

\subsection{Resolution Considerations}\label{sec:res}

The main limitation of this work is the relatively low resolution we 
have used,  $512^3$, which can affect our results in different
ways.  Numerical diffusivity has consequences on the 
thermal behavior of the simulations because it acts as a no controlled
thermal conductivity. Field's (1965) analysis shows that in the absence of 
thermal conductivity the linear growth rate of TI reaches its maximum value
and becomes constant for small enough scales, while for a thermally conducting
gas if the perturbation scale is reduced, the growth rate  
decreases to zero. The scale at which TI is suppressed by conduction 
$\lambda_{\rm F}$, is known as the Field length (Begelman \& McKee 1990). 
The effect of a small resolution is to impose a  ''numerical Field length''
which suppress the development of TI at artificially large scales, creating
broader boundaries between cool and warm gas and larger cold
structures. This effect is more relevant at low Mach numbers, when the 
pure development of TI dominates the simulations.
For 2-dimensional simulations with $512^2$ grid points, 
Gazol et al. (2005) have verified that density perturbations of 
amplitude 2.5\% and wavelength $\lambda = 16$ pixels 
(3.1 pc)  remain stationary (thus being the numerical Field length), 
while perturbations with $\lambda = 4$ pixels (0.8 pc) are completely 
damped in times ~3 Myr. They also explored the growth rate concluding
that $512^2$ is an acceptable resolution for capturing the linear growth, 
in the presence of realistic conductivity, of modes with sizes down to 
1/16 the box size. For the nonlinear case, density perturbations with 
large amplitude  velocity perturbations, the difference between 
the growth rates for $N = 512^2$ and $N = 1024^2$ is smaller than for 
the pure density perturbation case.
On the other hand the minimum size in our simulations ($\approx 0.2 $pc)
is larger than the typical size of the cold structures generated by 
pure TI ($\approx 0.1 $pc), thus our simulations probably overestimate 
the sizes of those cloudlets that are formed by the instability rather 
than by larger scale, coherent turbulent compressions, and certainly do 
not resolve their internal structure. As a consequence, improving the 
resolution could affect the results concerning density spectra at specific
temperature ranges because the cold and unstable gas distribution could
be modified. 

\begin{figure}
\epsscale{1.}
%\plotone{esp_ro_res.eps}
\plottwo{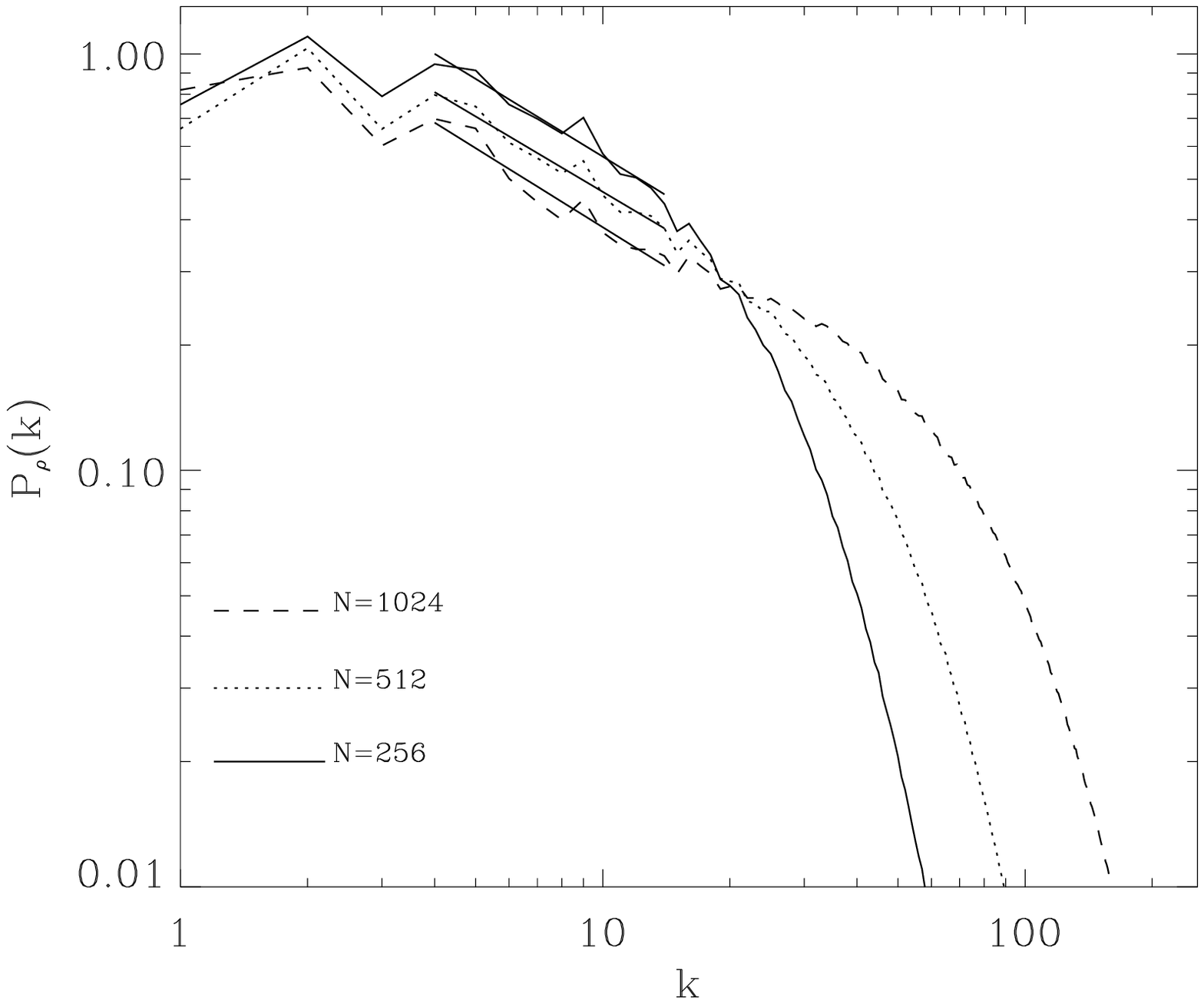}{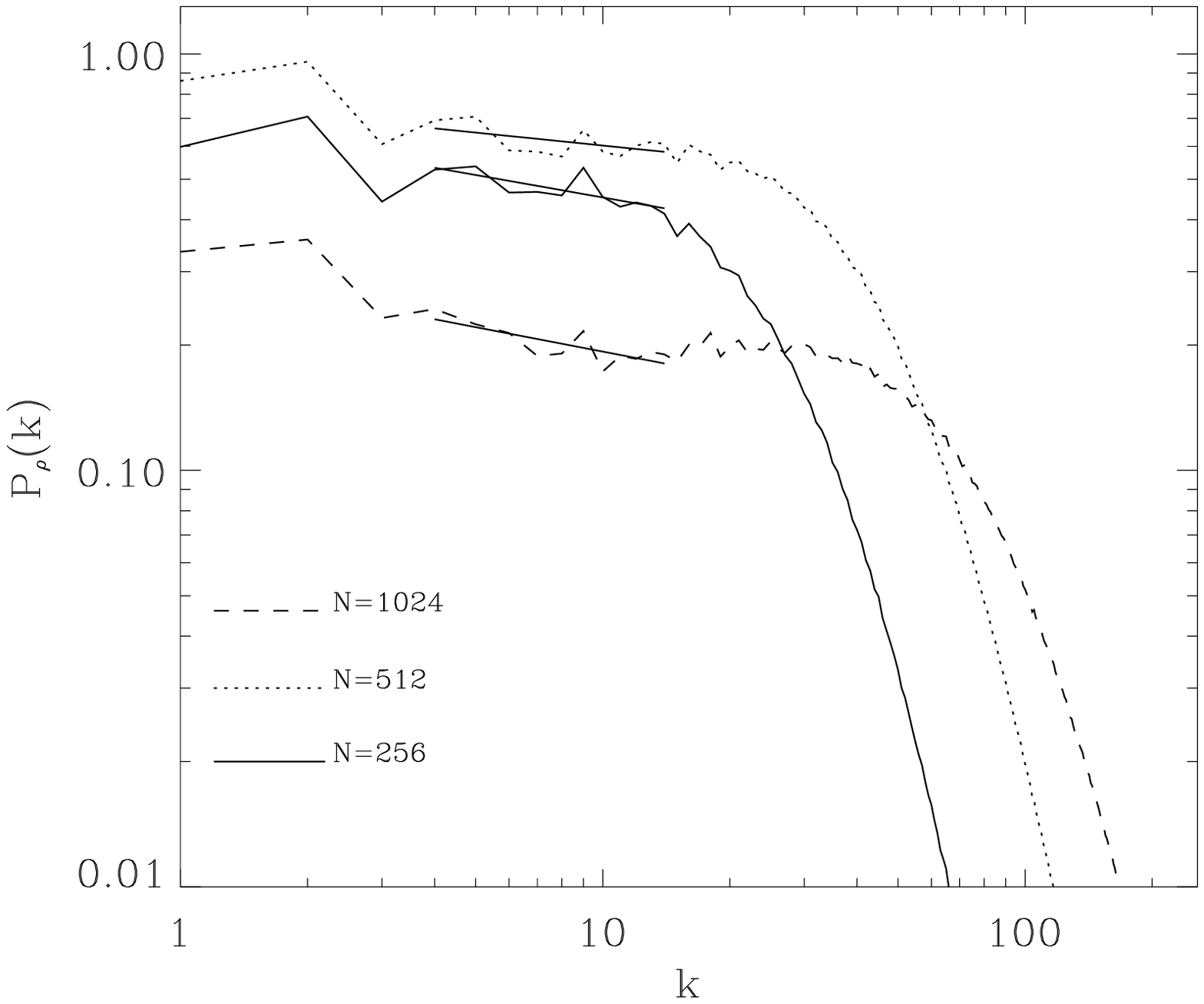}
\caption{Density power spectra $P_{\rho}$ for
simulations with $M=0.6$ ({\it left}) and $M=4.5$ ({\it right}), and 
 different resolutions.
Solid lines represent least-squares fits over the
range $4 \leq k \leq 14$.}
\label{fig:espectros_res}
\end{figure}

\begin{figure}
\epsscale{1.}
%\plotone{esp_v_res.eps}
\plottwo{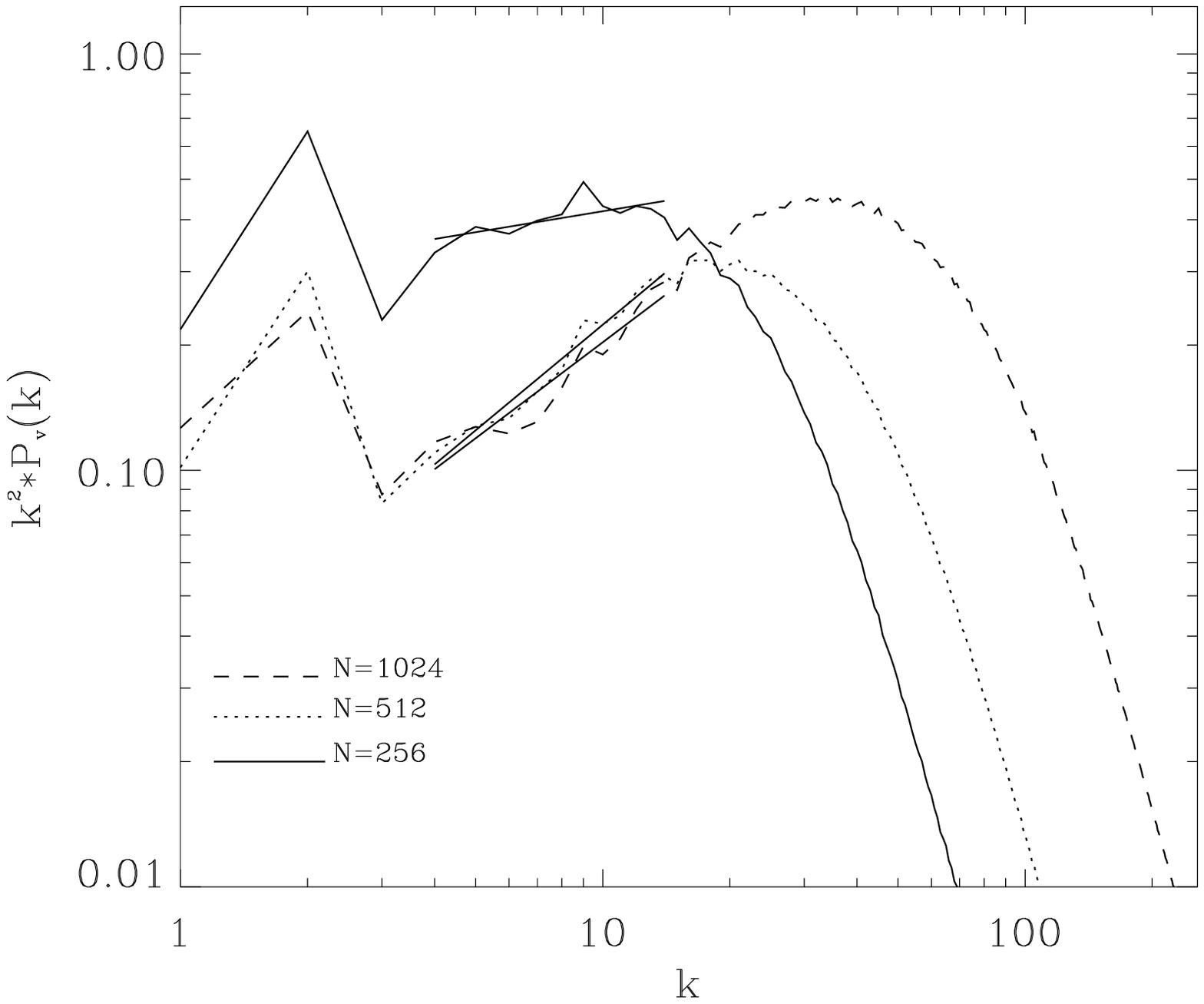}{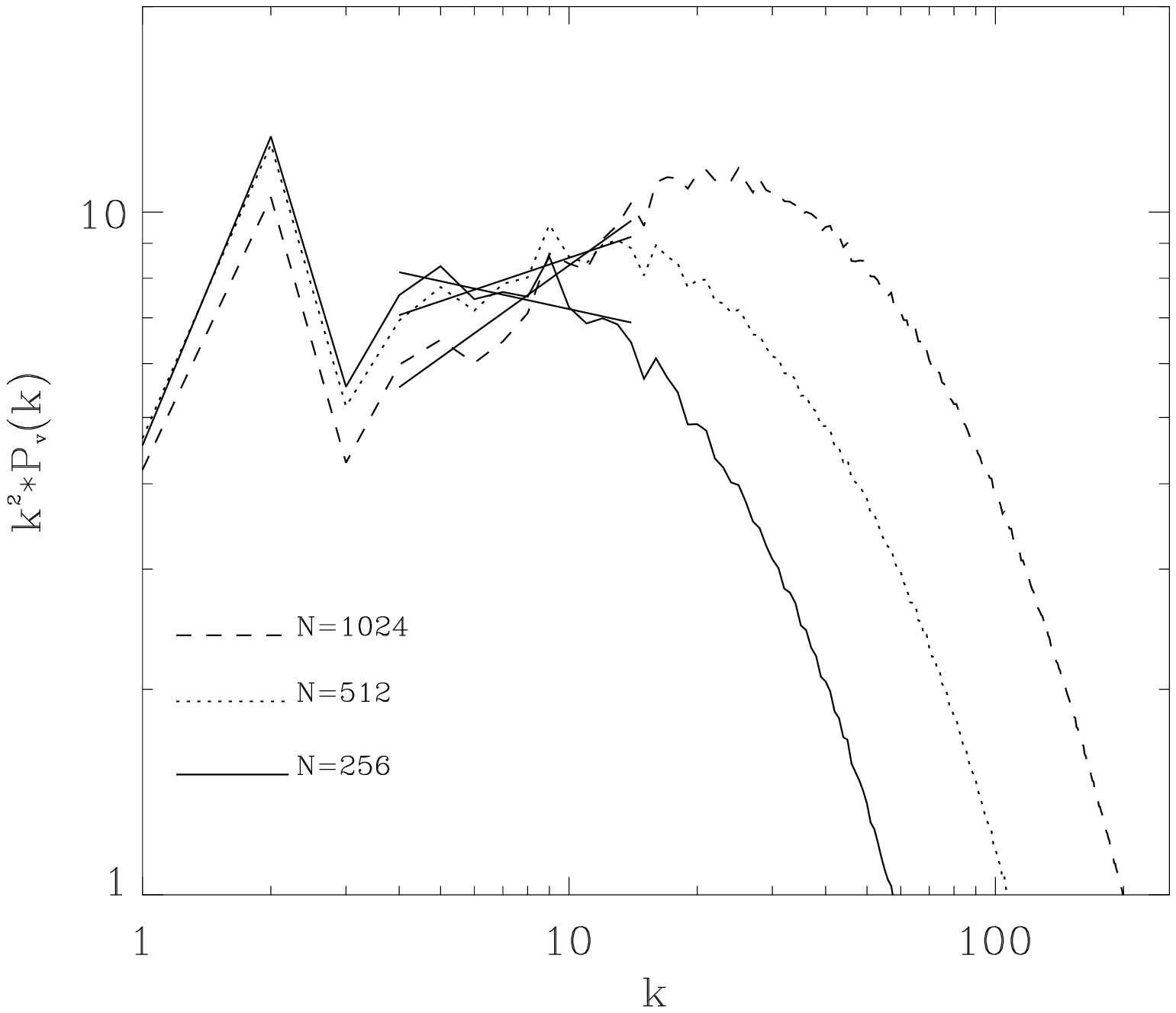}
\caption{Compensated velocity power spectra $k^2 P_{\bf v}$ for
simulations with  $M=0.6$ ({\it left}) and $M=4.5$ ({\it right}), 
and different resolutions.
Solid lines represent least-squares fits over the
range $4 \leq k \leq 14$.}
\label{fig:espectrosv_res}
\end{figure}

In Figure \ref{fig:espectros_res} ({\it left}) we show $P_{\rho}(k)$ for three 
simulations with $M=0.6$ and three different resolutions: $256^3$, $512^3$, 
and $1024^3$. As expected due to the increased numerical diffusivity, 
for the low resolution simulation the inertial range is very
short and worse defined than in the other cases. However, the slope of
the three spectra is very similar: -0.61,-0.60,-0.63. 
%A13
In the {\it right} panel of the same figure $P_{\rho}(k)$ is shown for
$M\approx 4.5$ ($M=$4.35, 4.55 and 4.51 for $1024^3$, $512^3$ and  $256^3$, 
respectively). In that case the effect of resolution is more important, 
slopes we find are -0.19, -0.10 and -0.25 for $1024^3$, $512^3$ and  $256^3$, 
respectively.
Velocity spectra for the same sets of simulations are displayed in Figure 
\ref{fig:espectrosv_res}. For $M=0.6$ ({\it left}) important differences 
can be appreciated between the $256^3$ spectrum on one hand and the $512^3$ 
and $1024^3$ on the other. For the last two cases slopes are -1.16 and -1.23 
at $512^3$ and $1024^3$, respectively. The {\it right} panel shows
$P_{v}(k)$ for $M\approx 4.5$, the resulting slopes in this case
are -2.13, -1.79, and -1.55  for $1024^3$, $512^3$ and  $256^3$, 
respectively. As for the density power spectrum, at larger $M$ values the 
effect of resolution is more important but the fact that we obtain an even
steeper spectrum for the large resolution simulation suggest that the
steepening of the velocity spectrum as $M$ increases is not a numerical
artifact. 
For the low resolution simulation
spectra have been computed by averaging over several snapshots at the
same time interval as the one stated in Table \ref{tab:tiempos} for the
$M=0.6$ simulation, while for the high resolution case we used few snapshots
around 4.5 turbulent crossing times. The resolution study on density spectra 
showed a better convergence
than that on velocity spectra.  It is due to the fact that, on density
fields, there is the additional constraint of  mass conservation inside
our computational box, irrespectively of the numerical resolution.

\section{Summary and Conclusions}\label{sec:conc}

In this paper we investigated the behavior of the density power spectrum of
forced, thermally bistable flows in physical conditions similar to those
of the atomic interstellar gas and with different Mach numbers. Our main
results cam be summarized as follows:

1. For thermally bistable flows, the density spectrum flattens as the Mach
number is increased. This is the same behavior as the one previously
reported for isothermal turbulence (Kim \& Ryu 2005), but  in the
present case the spectral slopes are shallower. We interpret this fact
and the behavior with $M$
as a consequence of a combined effect of the presence of strong
density fluctuations produced by TI and turbulence with 
the fact that the larger is $M$, the weaker are pressure forces in 
diffuse gas produced by the interaction of turbulence and TI.

2. The velocity power spectrum exhibits an opposite behavior, i.e. it becomes
steeper as the Mach number increases. The spectral indexes go from  values
much lower than the Kolmogorov slope for small $M$ to a value
larger than it for the larger Mach number. We hypothesize that the flattening
of the spectrum for low $M$ could be due to the increased relevance of 
the velocities generated by the development of TI, but a detailed
analysis of the velocity power spectrum for thermally bistable flows
is needed.

3. Comparison with observations show that the spectral slopes 
we obtain are consistent with some of the results on 
the three dimensional atomic power spectrum. If we use predictions
of the VCA method
to compare with observational slopes of the spatial power spectrum 
(which contains information about the three dimensional density as well
as about the three dimensional velocity field), then our results are 
also consistent with observations.

4. The density power spectrum at specific temperature ranges shows a behavior
that seems to be consistent with the first point of this summary but an
analysis with much larger numerical resolution is still needed.  

%A13
5. Resolution effects, in particular for large $M$ simulations, can
probably change the spectral slopes  but the behaviors we describe
and summarize in previous points are not modified by these effects.
 
In conclusion, the presence of TI does significantly affect the statistics
of turbulence. Different turbulence conditions, in particular different
Mach numbers, can lead to considerably different slopes of the density and
the velocity power 
spectrum of the atomic interstellar gas. Given the fact that the amount 
of available kinetic energy can change from one place to another, differences
in observational results concerning the density power spectrum could be, 
at least partially, be explained by this 
effect. Note however that in order to relate the density power spectrum
with the Mach number observational studies must clearly separate contributions
from the density distribution and contributions from the velocity field.

\acknowledgements 
We would like to acknowledge an anonymous referee for suggestions
to improve this work.
The work of A. G. was partially supported by UNAM-DGAPA grant IN111006-3.
The work of J.K. was supported by the National Research Foundation of
Korea through 2009-0062863 (ARCSEC). Numerical simulations were performed 
at the cluster Platform 4000 (KanBalam) at DGSCA, UNAM and at the Linux Cluster
for Astronomical Computations of KASI-ARCSEC.

\epsscale{1.0}

\epsscale{1.0}

\begin{deluxetable}{rrrrr}
\tablecolumns{5}
\tablewidth{0pc} 
\tablecaption{Time Parameters\label{tab:tiempos}}
\tablehead{
\colhead{$M$} & \colhead{$t_{\rm f}$\tablenotemark{a}} & \colhead{$\Delta t$\tablenotemark{b}} & \colhead{$N_{\rm outputs}$\tablenotemark{c}} &
\colhead{$t_{\rm f0}$\tablenotemark{d}} }
\startdata
0.2 & 6.0  & 2.0-6.0 & 21  & 30  \\ 
0.6 & 10.2 & 3.0-9.0 & 21  & 15  \\
1.3 & 9.0  & 2.8-8.3 & 24  & 7   \\
4.0 & 16.0 & 8.0-16.0 & 21 & 4   \\
4.5 & 18.0 & 9.0-18.0 & 21 & 4   \\
\enddata 
\tablenotetext{a}{End time of each simulation in units of the turbulent crossing time} 
\tablenotetext{b}{Time interval over which time averaged spectra are computed}
\tablenotetext{c}{Number of outputs used for temporal averages}
\tablenotetext{d}{End time of each simulation in code time  units} 
\end{deluxetable}

\begin{deluxetable}{rrrrrrrrr}
\tablecolumns{9}
\tablewidth{0pc} 
\tablecaption{Spectral Index \label{tab:indices}} 
\tablehead{ 
\colhead{$M$} & \colhead{$P_{\rho}$} & \colhead{$P_{\Sigma}$} & \colhead{$P_{\bf v}$} & \colhead{$P_{{\rho}^{1/2}\bf v}$} & \colhead{$P_{ {\rho}^{1/3}\bf v}$}& \colhead{$P_{w\rho}$\tablenotemark{a}} & \colhead{$P_{u\rho}$\tablenotemark{b}} & \colhead{$P_{c\rho}$\tablenotemark{c}}}
\startdata
0.2 & -0.84 & -1.64 & -1.08 & -0.77 & -0.89 & -0.77 & -0.79 & -0.84 \\ 
0.6 & -0.60 & -1.52 & -1.16 & -0.77 & -0.94 & -0.57 & -0.55 & -0.59 \\
1.3 & -0.20 & -1.14 & -1.51 & -0.80 & -1.11 & -0.51 & -0.27 & -0.12 \\
4.0 & -0.15 & -1.11 & -1.78 & -0.85 & -1.19 & -0.19 & -0.24 &  0.04 \\
4.5 & -0.10 & -1.19 & -1.79 & -0.84 & -1.18 & -0.18 & -0.19 &  0.09 \\
\enddata
 \tablenotetext{a}{Density power spectrum for gas with $T>6110$K}
 \tablenotetext{b}{Density power spectrum for gas with $313{\rm K}<T<6110$K}
 \tablenotetext{c}{Density power spectrum for gas with $T<313$K}
\end{deluxetable}

\end{document}